\documentclass[a4paper,twoside]{article}

\usepackage{epsfig}
\usepackage{subcaption}
\usepackage{calc}
\usepackage{amssymb}
\usepackage{amstext}
\usepackage{amsmath}
\usepackage{amsthm}
\usepackage{multicol}
\usepackage{pslatex}
\usepackage{apalike}
\usepackage{algorithm2e}
\usepackage[bottom]{footmisc}
\usepackage{multirow}
\usepackage[colorlinks]{hyperref}
\usepackage{SCITEPRESS}     %

\usepackage{booktabs}
\usepackage{cleveref}
\usepackage{siunitx}
\usepackage[inline]{enumitem}

\begin{document}

\title{Uncertainty Estimation for Super-Resolution using ESRGAN}

\author{\authorname{Maniraj Sai Adapa \sup{1}, Marco Zullich\sup{1}\orcidAuthor{0000-0002-9920-9095} and Matias Valdenegro-Toro\sup{1}\orcidAuthor{0000-0001-5793-9498}}
\affiliation{\sup{1}Department of AI, University of Groningen, Nijenborgh 9, 9747AG, Groningen, Netherlands}
\email{manirajadapa@gmail.com, m.zullich, m.a.valdenegro.toro@rug.nl}
}

\keywords{Super Resolution, Uncertainty Estimation, Computer Vision}

\abstract{Deep Learning–based image super-resolution (SR) has been gaining traction with the ad of Generative Adversarial Networks. Models like SRGAN and ESRGAN are constantly ranked between the best image SR tools. However, they lack principled ways for estimating predictive uncertainty. In the present work, we enhance these models using Monte Carlo–Dropout and Deep Ensemble, allowing the computation of predictive uncertainty. When coupled with a prediction, uncertainty estimates can provide more information to the model users, highlighting pixels where the SR output might be uncertain, hence potentially inaccurate, if these estimates were to be reliable. Our findings suggest that these uncertainty estimates are decently calibrated and can hence fulfill this goal, while providing no performance drop with respect to the corresponding models without uncertainty estimation.}

\onecolumn \maketitle \normalsize \setcounter{footnote}{0} \vfill

\section{\uppercase{Introduction}}
\label{sec:intro}

\begin{figure*}[t]
    \centering
    \begin{subfigure}{0.14\textwidth}
        \includegraphics[width=\linewidth]{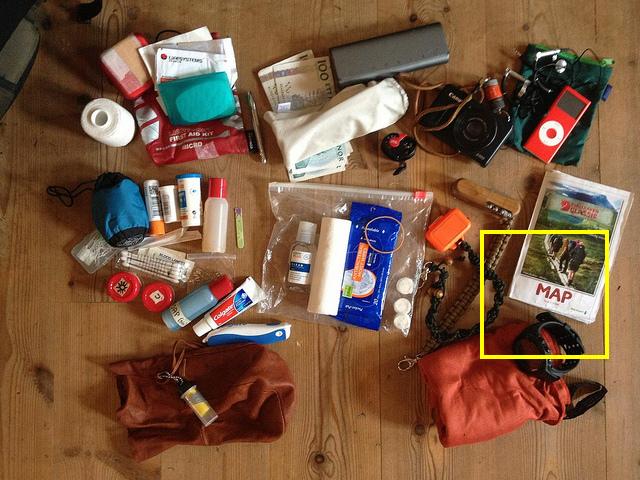}
        \caption*{Image}
    \end{subfigure}    
    \begin{subfigure}{0.14\textwidth}
        \includegraphics[width=\linewidth]{figures/gfx_en/cluuter_480_230/gen_img.png}
        \caption*{SR}
    \end{subfigure}    
    \begin{subfigure}{0.14\textwidth}
        \includegraphics[width=\linewidth]{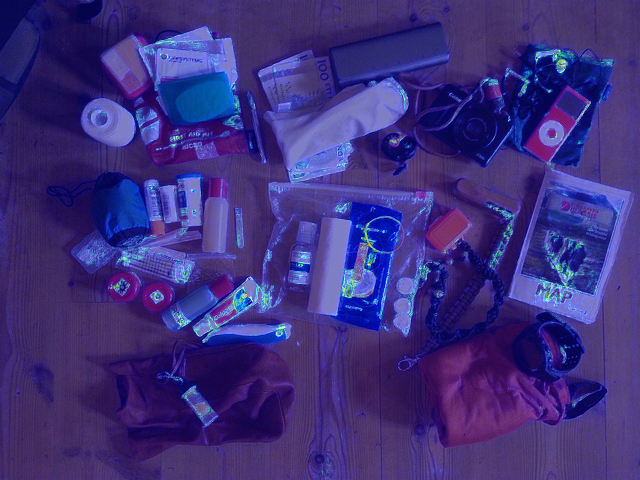}
        \caption*{Uncert Overlay}
    \end{subfigure}
    \begin{subfigure}{0.14\textwidth}
        \includegraphics[width=\linewidth]{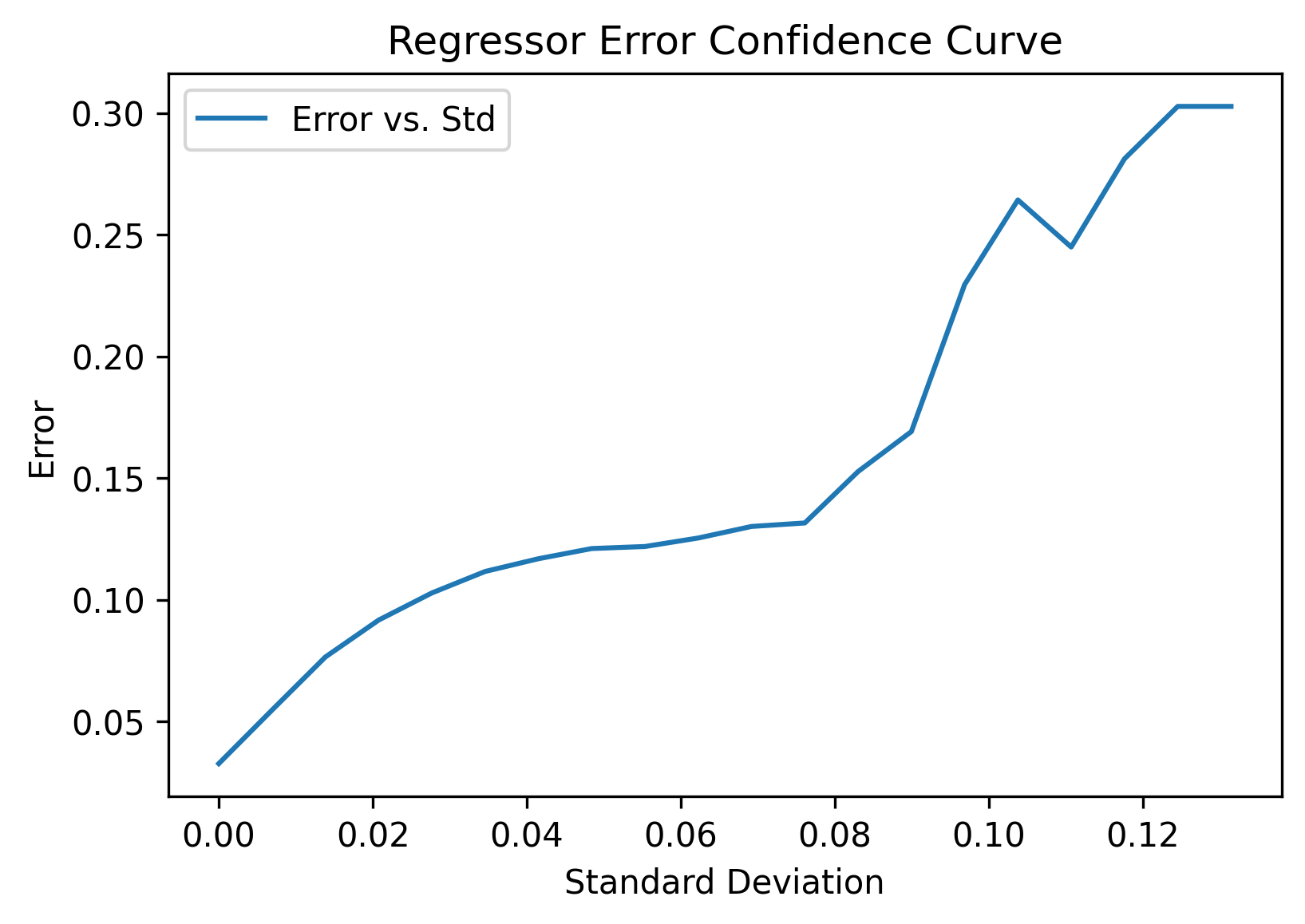} 
        \caption*{Error vs Std}
    \end{subfigure}        
    \begin{subfigure}{0.10\textwidth}
        \includegraphics[width=\linewidth]{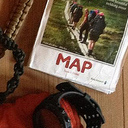}
        \caption*{Crop}
    \end{subfigure}
    \begin{subfigure}{0.10\textwidth}
        \includegraphics[width=\linewidth]{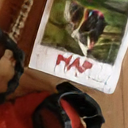}
        \caption*{SR Crop}
    \end{subfigure}
    \begin{subfigure}{0.10\textwidth}
        \includegraphics[width=\linewidth]{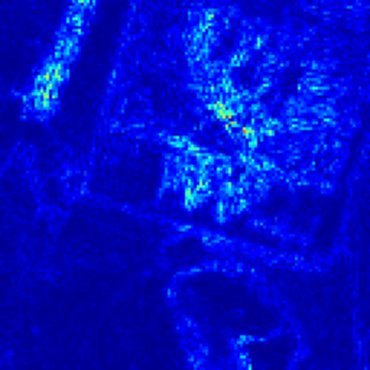}
        \caption*{Uncert}
    \end{subfigure}

    \vspace*{1em}
    \begin{subfigure}{0.14\textwidth}
        \includegraphics[width=\linewidth]{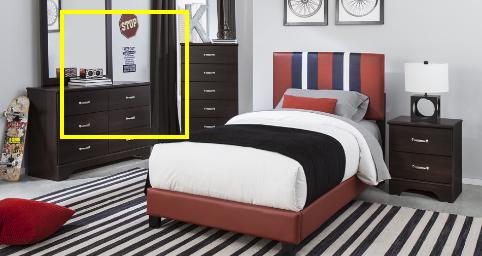}
        \caption*{Image}
    \end{subfigure}    
    \begin{subfigure}{0.14\textwidth}
        \includegraphics[width=\linewidth]{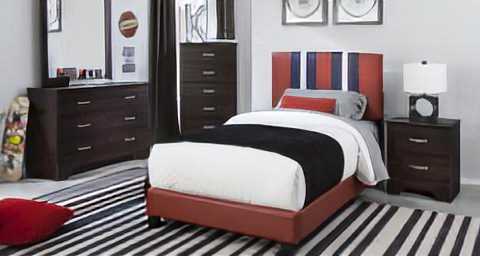}
        \caption*{SR}
    \end{subfigure}    
    \begin{subfigure}{0.14\textwidth}
        \includegraphics[width=\linewidth]{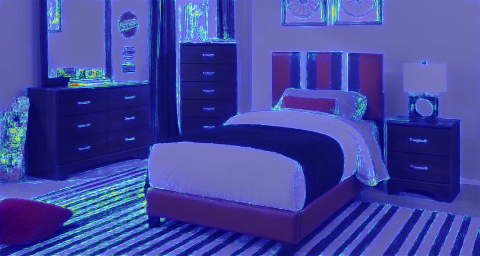}
        \caption*{Uncert Overlay}
    \end{subfigure}
    \begin{subfigure}{0.14\textwidth}
        \includegraphics[width=\linewidth]{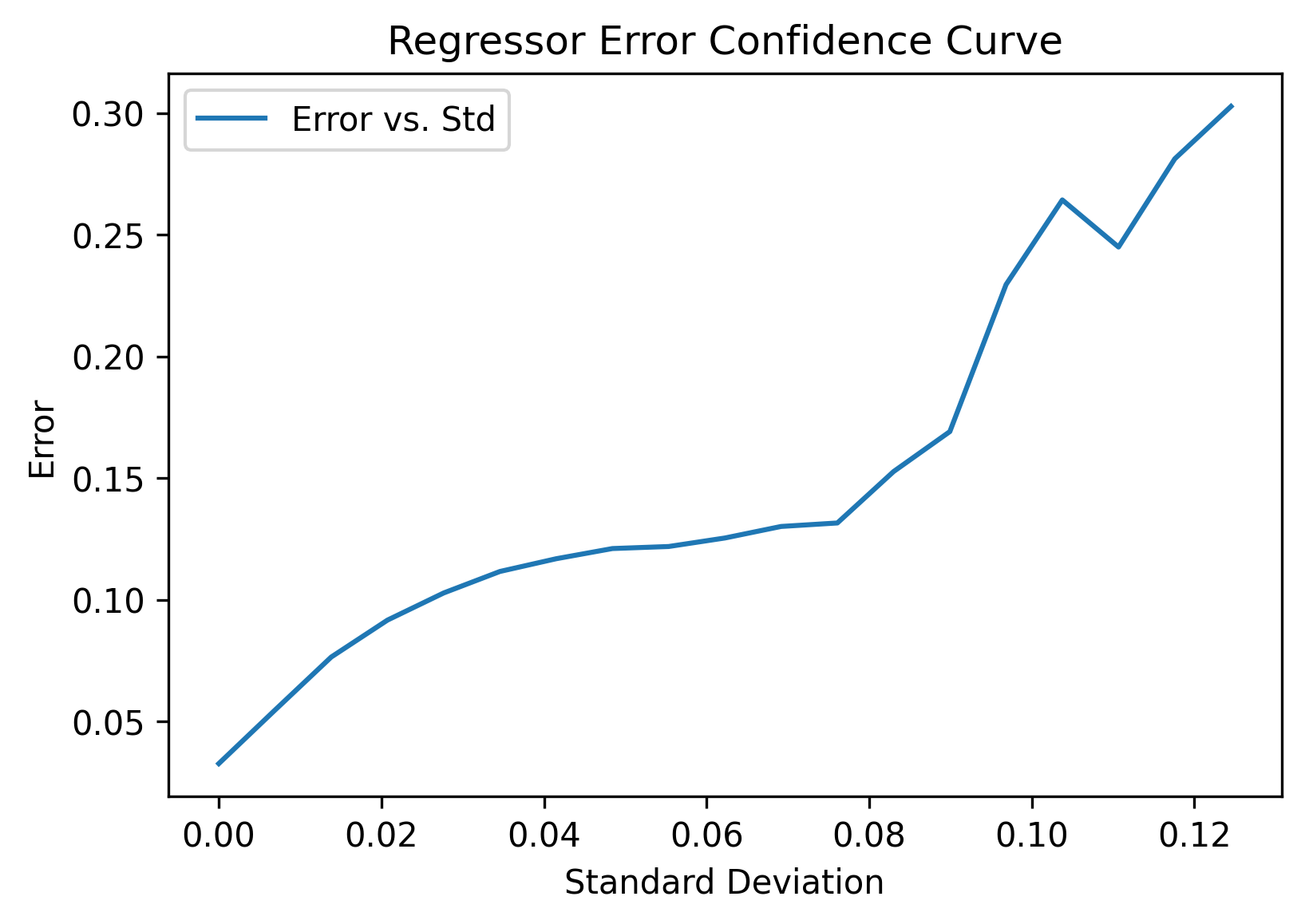} 
        \caption*{Error vs Std}
    \end{subfigure}
    \begin{subfigure}{0.10\textwidth}
        \includegraphics[width=\linewidth]{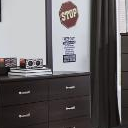}
        \caption*{Crop}
    \end{subfigure}
    \begin{subfigure}{0.10\textwidth}
        \includegraphics[width=\linewidth]{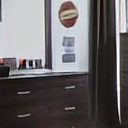}
        \caption*{SR Crop}
    \end{subfigure}
    \begin{subfigure}{0.10\textwidth}
        \includegraphics[width=\linewidth]{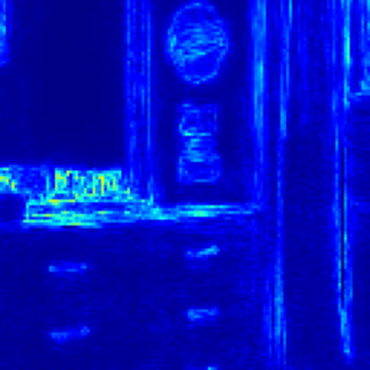}
        \caption*{Uncert}
    \end{subfigure}

    \vspace*{1em}
    \begin{subfigure}{0.14\textwidth}
        \includegraphics[width=\linewidth]{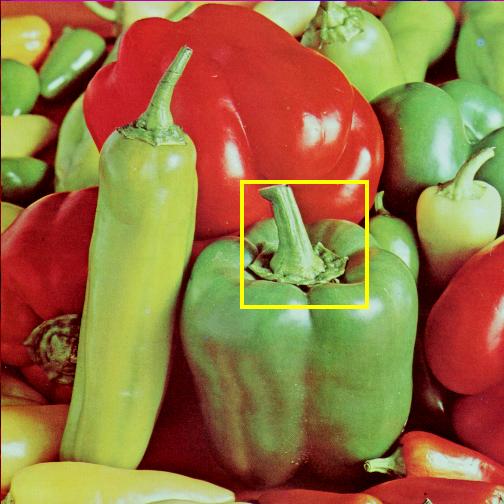}
        \caption*{Image}
    \end{subfigure}    
    \begin{subfigure}{0.14\textwidth}
        \includegraphics[width=\linewidth]{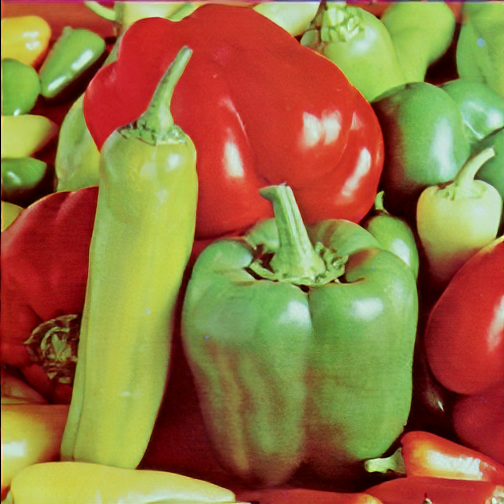}
        \caption*{SR}
    \end{subfigure}    
    \begin{subfigure}{0.14\textwidth}
        \includegraphics[width=\linewidth]{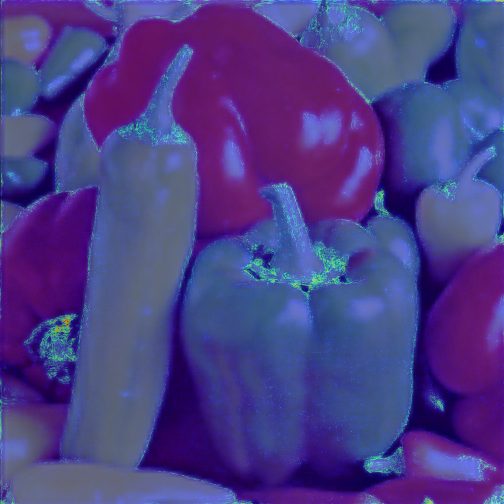}
        \caption*{Uncert Overlay}
    \end{subfigure}
    \begin{subfigure}{0.14\textwidth}
        \includegraphics[width=\linewidth]{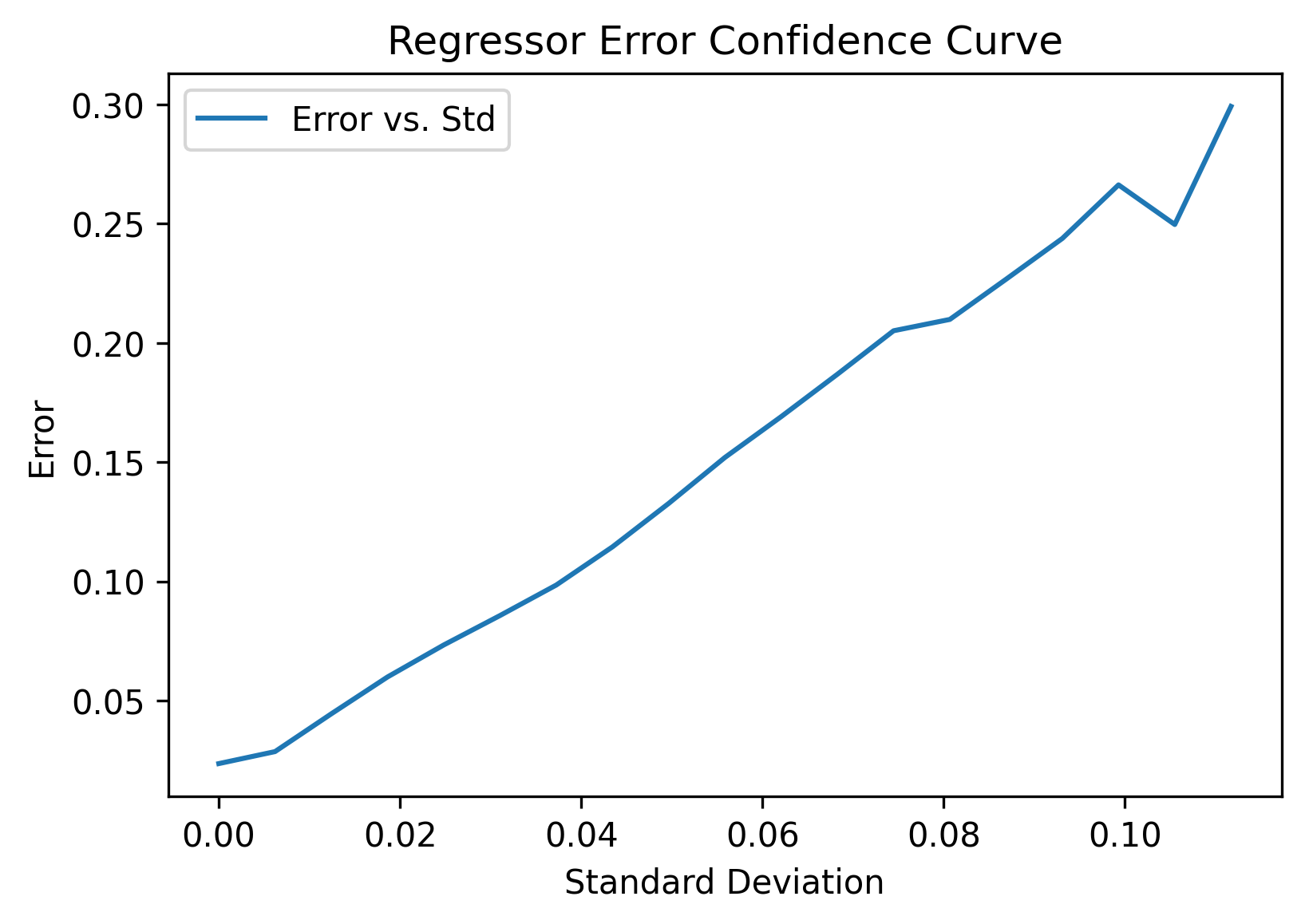} 
        \caption*{Error vs Std}
    \end{subfigure}    
    \begin{subfigure}{0.10\textwidth}
        \includegraphics[width=\linewidth]{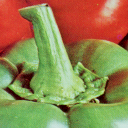}
        \caption*{Crop}
    \end{subfigure}
    \begin{subfigure}{0.10\textwidth}
        \includegraphics[width=\linewidth]{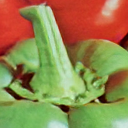}
        \caption*{SR Crop}
    \end{subfigure}
    \begin{subfigure}{0.10\textwidth}
        \includegraphics[width=\linewidth]{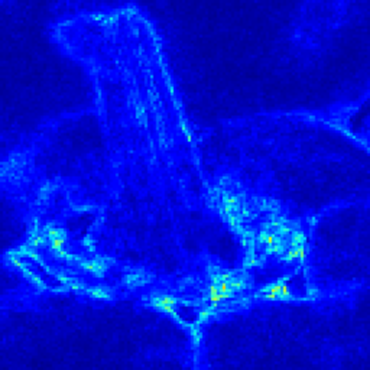}
        \caption*{Uncert}
    \end{subfigure}

    \caption{Ensemble Results using ESRGAN with Uncertainty, including error vs standard deviation plots. This figure shows how SR uncertainty correlates with SR reconstruction errors and can be used to detect possible errors at inference time. Error vs Std plots show that uncertainty correlates very well with absolute errors at the pixel level.}
    \label{fig:ens_esrgan_visual}
\end{figure*}

Super-Resolution (SR) is an important computer vision task, where a low-resolution image is upscaled to a higher resolution one. It is fundamentally an inverse problem, where missing information needs to be filled by making assumptions encoded in a model, which can lead to errors, as shown in Figure \ref{fig:ens_esrgan_visual}.

Many efforts are made to improve SR models to increase their accuracy, but any model will tend to produce erroneous outputs if the input is outside the training distribution. An important task is then provide feedback to a human user on which pixels or regions of the SR output image are likely to be incorrect or imprecise.

In this paper we combine two uncertainty estimation methods with a state of the art SR model---Super Resolution Generative Adversarial Network (SRGAN) \cite{ledig2017photo} and Enhanced SRGAN (ESRGAN) \cite{wang2018esrgan}---, to build a SR model with epistemic uncertainty estimation, which outputs a SR image and a uncertainty map, indicating which regions are likely to be incorrect. We evaluate the performance of uncertainty estimation, noting that an ensemble of 5 ESRGAN generators works the best, and provide extensive quantitative and qualitative results, showcasing the usefulness of uncertainty estimation in the SR domain.

We posit our results show that uncertainty estimation, in particular ensembles, can provide useful feedback to a human using SR results, and the standard deviation produced by the model can work as a per-pixel error proxy.

The contributions of this paper are: we build SRGAN and ESRGAN models with uncertainty estimation, we evaluate uncertainty performance in several well known datasets, and validate that model standard deviation can be used as a proxy for test time error.

This work expands the state of the art by building simple combinations of a state of the art SR model (ESRGAN) with Monte Carlo Dropout and Ensembles, with an explicit focus on qualitative and quantitative uncertainty estimation, showing that uncertainty can be used as a proxy for error at inference time, for natural color images. Note that our work is not about improving the super-resolution task performance, but we argue that fundamentally any super-resolution model will make mistakes at some point, especially with out of distribution images, and uncertainty estimation is a key component to notify the end user about these mistakes via higher per-pixel uncertainty.

\section{State of the Art}

The literature for super resolution uncertainty is relatively underexplored. Kar and Biswas \cite{kar2021fast} use stochastic Batch Normalization for uncertainty estimation in deep SISR models. Liu et al. \cite{liu2023spectral} estimate uncertainty in spectral domain instead of the common spatial domain for the DDL-EDSR model.

SR Uncertainty outside of natural images is also present. Tanno et al \cite{tanno2017bayesian} use variational dropout for SR of 3D Diffusion MRI brain images, while Song and Yang \cite{song2023uncertainty} use Bayesian Neural Networks for SR in wave array imaging and making separate predictions of aleatoric and epistemic uncertainty.

Most previous research on SR uncertainty focuses on improving SR accuracy using uncertainty estimation (\cite{kar2021fast} and \cite{liu2023spectral}), or are applied to domains outside of natural images. There is often not a deep focus on uncertainty quantification for SR and its consequences.

We perform a deep evaluation of uncertainty quality for SR, as we assume that SR models will always make errors in out of distribution settings, and per-pixel output uncertainty can guide the human user to detect these errors.

\section{\uppercase{ESRGAN with Uncertainty}}

\subsection{Image SR Using GAN-based Models}

SRGAN \cite{ledig2017photo} and ESRGAN \cite{wang2018esrgan} have emerged as industry-standard architectures for image super-resolution. 
Both techniques make use of a Generative Adversarial Network (GAN) framework, whereas a \emph{generator} is trained to super-resolve low-resolution images, while a \emph{discriminator} is simultaneously trained to distinguish between real high-resolution images and the output of the generator.
SRGAN demonstrates the potential of GANs for super-resolution by utilizing an adversarial loss versus the discriminator. Building upon these foundations, ESRGAN further enhances the approach with improvements in both generator and discriminator architecture, focusing on optimizing perceptual quality. 
The success of SRGAN and ESRGAN has made them widely adopted baseline approaches, with code and pre-trained models readily available. While these pre-trained GAN models offer strong performance, training customized models from scratch can provide advantages when exploring specific techniques like uncertainty estimation and also provide architectural similarity.

Formally, we call $D$ the discriminator and $G$ the generator.
Each training data point is composed by a high-resolution image $Y\in\mathbb{R}^{h\times w}$ acting as ground truth and its low-resolution equivalent $X\in\mathbb{R}^{h^\prime \times w^\prime}$, with $h>h^\prime, w>w^\prime$, acting as input.
Depending on the specific dataset employed, $X$ is usually obtained from $Y$ by applying a specific downsampling technique, such as bicubic interpolation.
The generator $G$ takes as input $X$ and produces a super-resolved image $\hat{Y}\in\mathbb{R}^{h \times w}$.
The discriminator $D$ is fed a high-resolution image (either $Y$ or $\hat{Y}$) and outputs a scalar $r\in (0,1)$, which can be interpreted as a \emph{probability} value of that image being a real high-resolution image.

\subsubsection{SRGAN} 
The discriminator of the SRGAN plays a crucial role in adversarial training. 
As the discriminator becomes effective in distinguishing the super-resolved images, the adversarial process forces the generator to produce images that are increasingly realistic.

The discriminator is a simple convolutional neural network classifier that consists of a series of strided convolution layers that are responsible for extracting hierarchical features and a final classification layer which produces the scalar output.

SRGAN uses a loss function for the generator $l_G^{\text{SRGAN}}$ for improving the perceptual quality of super-resolution results.
It is composed of two objectives:

\begin{equation}
    l^{\text{SRGAN}}_G = \underbrace{l_{\text{perc}}^{\text{SRGAN}}}_\text{perceptual loss} + 10^{-3}\cdot\underbrace{l_{\text{adv}}^{\text{SRGAN}}}_\text{adversarial loss}.
\end{equation}

The perceptual loss is computed on a VGG19 \cite{simonyan2014very} network which is pre-trained on Imagenet\cite{deng2009imagenet}.
The goal of this component is to enforce a realistic output which can be identified as \emph{realistic} by the VGG19 model.
It is defined as the Euclidean distance between the feature representations from deeper layers of this model for the super-resolved image and the original high-resolution image was termed the content loss.
If we call the backbone of the VGG19 model $f_{\text{VGG}}$, the perceptual loss is:

\begin{equation}\label{eq:contentloss}
    l^{\text{SRGAN}}_{\text{perc}}(X, Y) = ||f_{\text{VGG}}(G(X)) - f_{\text{VGG}}(Y)||^2_2.
\end{equation}

The intuition is that $f_{\text{VGG}}$ will project plausible generator outputs close to the embedding of the corresponding ground truth image.

The adversarial loss is inspired by the original GAN loss \cite{goodfellow2014generative}, which is based off of the binary cross entropy loss on the discriminator output.
In this case, it acts only on the super-resolved images, disregarding the real ones:

\begin{equation}
    l^{\text{SRGAN}}_{\text{adv}}(X) = - \log D_{\theta_D} (G_{\theta_G} (X)).
\end{equation}

The discriminator loss is, instead, the same proposed by Goodfellow et al. \cite{goodfellow2014generative} in the original GAN paper.

\subsubsection{ESRGAN}
ESRGAN builds upon SRGAN, implementing several enhancements to the generator and discriminator.
Since its introduction, ESRGAN has consistently achieved state-of-the-art results on standard benchmarks and is regarded as one of the top-performing single-image super-resolution methods.

The main improvement is the introduction of residual-in-residual dense blocks (RRDBs) in the generator to extract more image details. Each RRDB consists of several Residual Dense Blocks (RDB). In an RDB, the output of each convolutional layer is concatenated with the inputs of all subsequent layers, promoting feature reuse and the learning of fine texture details.

The features learned through each RRDB are aggregated, and a global residual learning connection is added to form the final high-dimensional feature maps. These feature maps are then up-scaled to higher resolution using pixel-shuffle \cite{shi2016real}.

The discriminator in ESRGAN follows the design of a standard GAN discriminator but with some modifications. The discriminator applies the principles of Relativistic GAN (RGAN) \cite{jolicoeur2018relativistic} for stabilizing the training.
RGANs improve on the GAN objective by comparing the likelihood of the super-resolved image relative to the real high-resolution counterpart.
In GANs, instead, the (absolute) likelihood of the real image is used as objective instead.

The loss function of ESRGAN  utilizes a weighted combination of three components to optimize the tradeoff between pixel-level accuracy and perceptual similarity.
These components are, respectively, the perceputal loss, the adversarial loss, and the perceptual loss:

\begin{equation}
    l^{\text{ESRGAN}} = \lambda_{\text{cont}}l_{\text{cont}}^{\text{ESRGAN}} + \lambda_{\text{adv}}l_{\text{adv}}^{\text{ESRGAN}} +
    \lambda_{\text{perc}}l_{\text{perc}}^{\text{ESRGAN}}
\end{equation}

The adversarial loss relies on the RGAN principle, thus comparing the discriminator behavior when evaluating real and super-resolved images to the corresponding super-resolved or real counterpart, respectively.

The content loss is now defined as a L1-norm reconstruction loss: 
\begin{equation}\label{eq:l1-reconstruction}
    l^{\text{ESRGAN}}_{\text{cont}}(X, Y)=|| G(X) - Y ||_1.
\end{equation}

The perceptual loss is modified from \Cref{eq:contentloss} by considering the L1 norm of the VGG19 embeddings instead of the L2 norm.

\subsection{Uncertainty Estimation for Super-Resolution}

In the current work, we make use of two popular techniques for uncertainty estimation, namely Monte-Carlo Dropout (MCD) and Deep Ensembles (DEs).

\subsubsection{Monte Carlo Dropout}

MCD \cite{gal2016dropout} is a framework for training approximate Bayesian Neural Networks by modifying the behavior of the regularization technique dropout \cite{srivastava2014dropout}.
While dropout randomly zeroes out, with a given probability $p_\text{drop}$, certain activations in a specific feature maps  during the training phase, the main intuition behind MCD is to keep this behavior active during inference, thus obtaining a stochastic output.
Samples from the posterior predictive distribution can be obtained by performing $M$ stochastic forward passes through the model with different randomly sampled dropout masks.
The standard deviation of the predictions can be computed as an estimate of uncertainty.
A major appeal of MCD is its ease of implementation---if the deterministic model is already equipped with dropout layers, no changes are needed to the underlying architecture or training process.
In the case of SR, given the $M$ outputs $\hat{Y}^{(1)},\dots,\hat{Y}^{(M)}$, we aggregate them into a mean output image:
\begin{multline}\label{eq:mean}
    \mu(\hat{Y})_{i,j} = \frac{1}{M} \sum_{m=1}^{M} \hat{Y}^{(m)}_{i,j},\\ i\in\{1,\dots,h\}, w\in\{1,\dots,w\}.
\end{multline}
With $\hat{Y}(X) = G(X)$ being one forward pass of the generator. Similarly, we can compute a per-pixel standard deviation:
\begin{multline}
    \label{eq:std}
    \sigma(\hat{Y})_{i,j} = \sqrt{\frac{1}{M^2} \sum_{m=1}^M (\hat{Y}^{(m)}_{i,j}-\mu(\hat{Y})_{i,j})^2},\\
    i\in\{1,\dots,h\}, w\in\{1,\dots,w\}.
\end{multline}

\subsubsection{Deep Ensembles}

DEs are composed of $M$ models with the same architecture, trained on the same dataset, but starting from different random initializations of the parameters.
At inference time, the predictions can be aggregated analogously to MCD, as shown in \Cref{eq:mean,eq:std} by taking into consideration that the number of samples is now equivalent to the number of components in the ensemble.
Despite not being approximate Bayesian Neural Networks---the output of the $M$ components are always deterministic---, DEs are often treated as the most reliable Deep Learning method for uncertainty estimation \cite{lakshminarayanan2017simple}.
This is especially true for the detection of Out-of-Distribution (OOD) data:
for familiar, in-distribution data, the $M$ components will likely agree on their prediction, while, for OOD data, the predictions will likely be random;
even in case of highly-confident single predictions, $\mu(\hat{Y})$ will be a high-entropy, smooth simplex.

\subsection{Data}

In the current work, we made use of several dataset for training and evaluating SRGAN and ESRGAN.
For training, we made use of a combination of the following datasets, which are common choices for SR tasks: DIV2K \cite{agustsson2017ntire} (\num{1000} images), UHDSR4K \cite{zhang2021benchmarking} (\num{5999} images in the training split), and 
Flickr2K \cite{timofte2017ntire} (\num{2650} images).
For evaluating the model, we make use of additional datasets:
Microsoft COCO \cite{lin2014microsoft}, Set5 \cite{bevilacqua2012low}, Set14 \cite{zeyde2012single}, Urban100 \cite{huang2015single}, and BSD100 \cite{martin2001database}.
The latter four are small-scale, high-resolution datasets. We use a selection of images for qualitative evaluation from COCO, BSD100 and Urban10, and we  perform quantitative evaluation on Set5 an Set14.

To enforce consistency in the image size, for all datasets, we cropped the images to a common resolution of $256\times 256$ px to obtain a ground truth image, while, to obtain the corresponding low-resolution input, we used bicubic interpolation downsampling with a target size of $64 \times 64$ px.
While usually the original data in these datasets is of much higher resolution (normally above \num{1000} px per side), we had to reduce this to the much more achievable $256 \times 256$ to limit the computational requirements of training and running inference using MCD and DEs.

\begin{table}[t]
    \centering
    \begin{tabular}{llllll}
        \toprule
        &				& \multicolumn{2}{c}{Set 5} & \multicolumn{2}{c}{Set 14} \\
        &  Uncert & PSNR  & SSIM & PSNR  & SSIM\\
        \midrule                       
        \multirow{3}{*}{\rotatebox{90}{\small SRGAN}}   & None 		& 29.10  & \textbf{0.8289} & 25.59 & 0.7232\\
        & MCD 		& 28.25 & 0.8117 & 24.72 & 0.7156\\ 
        & Ensemb 	& \textbf{29.25} & 0.8244 & \textbf{25.81} & 0.\textbf{7245}\\
        \midrule
        \multirow{3}{*}{\rotatebox{90}{\small ESRGAN}}  & None 		& 32.02 & 0.8923 & 27.11 & \textbf{0.7784}\\ 
        & MCD		& 32.23 & 0.8972 & 26.85 & 0.7611\\         
        & Ensemb	& \textbf{32.68} & \textbf{0.8997} & \textbf{27.23} & 0.7692\\         
        \bottomrule
    \end{tabular}
    \caption{Comparison of Baseline and Uncertainty Estimation Techniques on Set 5 and Set 14 datasets.}
    \label{tab:uncert_num_comparison}
\end{table}

\section{\uppercase{Experiments}}

\subsection{Model and Uncertainty Evaluation}
We operated the evaluation of the SR quality according to two popular metrics, Peak Signal-to-Noise Ratio (PSNR) and Structural Similarity Index (SSIM).
The former evaluates the reconstruction quality, while the latter is an attempt at computing a perceptual similarity based upon structural information, luminance, and contrast.
Both of these metrics have their limitations ecc ecc...

For what concerns the evaluation of the uncertainty estimates, we make use of error vs.~standard deviation plots.
Considering a test dataset of $n$ units, we compute the mean standard deviation across each of these images.
Given a generic image $k\in\{1,\dots,n\}$, $\tilde{\sigma}^{(k)} = \frac{1}{h\cdot w}\sum_{i,j} \sigma^{(k)}_{i,j},$ where $i,j$ represent generic pixels of this image.

We can proceed to bin the various $\tilde{\sigma}$'s in the test set, calculating a corresponding error metric for each of the images in the bin.
The underlying idea is that uncertain predictions (i.e., predictions with a high standard deviation) should have, on average, a high error, while confident predictions (with low standard deviation) should have a corresponding low error.
This can be visualized in a chart, whereas, by plotting the various values of error and standard deviation, a clear linear ascending trend should be visible.

In our specific case, we decided to use the per-pixel Mean Absolute Error (MAE) between the ground truth and super-resolved image as error metric.

In addition to the quantitative evaluation, we provide a qualitative assessment, both of the model performance and the uncertainty estimation.

\subsection{Experimental Setup}

\subsubsection{SRGAN}
For SRGAN, we trained the generator and discriminator networks from scratch using the Kaiming normal initialization \cite{he2015delving} for the convolutional layers. 
We used the Adam optimizer \cite{kingma2014adam} with momentum terms $\beta_1$=0.9 and $\beta _2$=0.999. 
We set the initial learning rate to $0.0001$.
We also employed data augmentation by means of 
\begin{enumerate*}[label=(\alph*)]
    \item random cropping,
    \item random rotation by an angle of 90°, 180°, or 270°,
    \item random horizontal flip, and
    \item random vertical flip.
\end{enumerate*}
We trained both the discriminator and the generator for a total of 300 epochs with a batch size of 16, for a total of approximately 40 hours of wall-time.

\subsubsection{ESRGAN}

The training process consisted of two phases. 
First, we pre-trained the generator on lower-resolution images using the L1 reconstruction loss from \Cref{eq:l1-reconstruction} to optimize Peak Signal-to-Noise Ratio (PSNR).
We initialized the learning rate at \num{0.0001}.

Next, we proceed to train alternatively the discriminator and generator for a total of 200 epochs, with a batch size of 16. 
For this phase, we use the Adam optimizer with learning rate \num{0.0001}, with decay factor of 2 after \numlist{25;50;100;150} epochs.
Analogously to SRGAN, we also employed data augmentation.

\subsubsection{MCD and DE}

In order to apply MCD for uncertainty estimation, we modified the baseline SRGAN and ESRGAN models by incorporating dropout layers, since these were not included in the original implementations.
We added 4 (for SRGAN) and 5 (for ESRGAN) dropout layers throughout the generator architecture with $p_\text{drop}=0.1$.
In order to reduce computational requirements, we opted for $M=10$ for MCD and $M=5$ for DEs.
For training the models, we used the same optimizers, hyperparameters, and data augmentation routines employed in the original models, as illustrated in the previous paragraphs.

\subsection{SRGAN vs ESRGAN SR Comparison}

\begin{figure*}[!ht]
    \centering        
    \begin{minipage}[t]{0.32\textwidth}
        \begin{subfigure}{0.35\linewidth}
            \includegraphics[width=\linewidth]{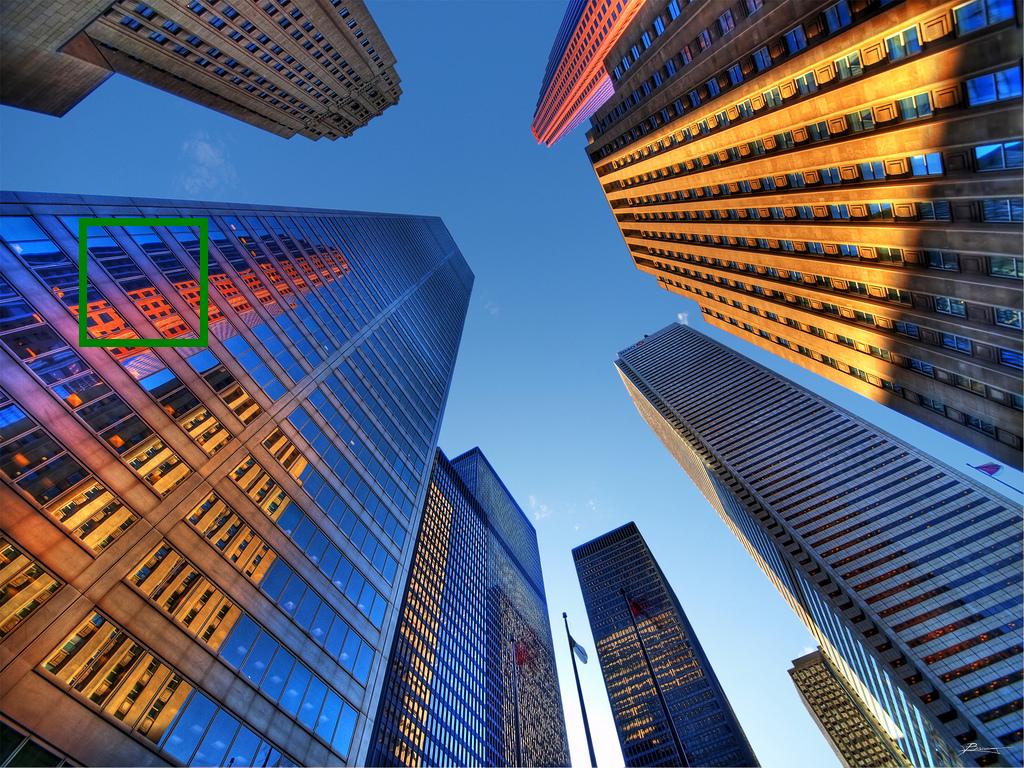}
            \caption*{Orig Image + Crop}
        \end{subfigure}
        \begin{subfigure}{0.20\linewidth}
            \includegraphics[width=\linewidth]{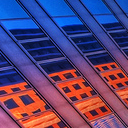}
            \caption*{HR\vspace*{1.1em}}
        \end{subfigure}            
        \begin{subfigure}{0.20\linewidth}
            \includegraphics[width=\linewidth]{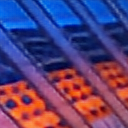}
            \caption*{SRGAN}
        \end{subfigure}        
        \begin{subfigure}{0.20\linewidth}
            \includegraphics[width=\linewidth]{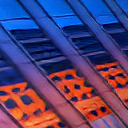}
            \caption*{ESRGAN}
        \end{subfigure}
    \end{minipage}
    \,
    \begin{minipage}[t]{0.32\textwidth}
        \begin{subfigure}{0.35\textwidth}
            \includegraphics[width=\linewidth]{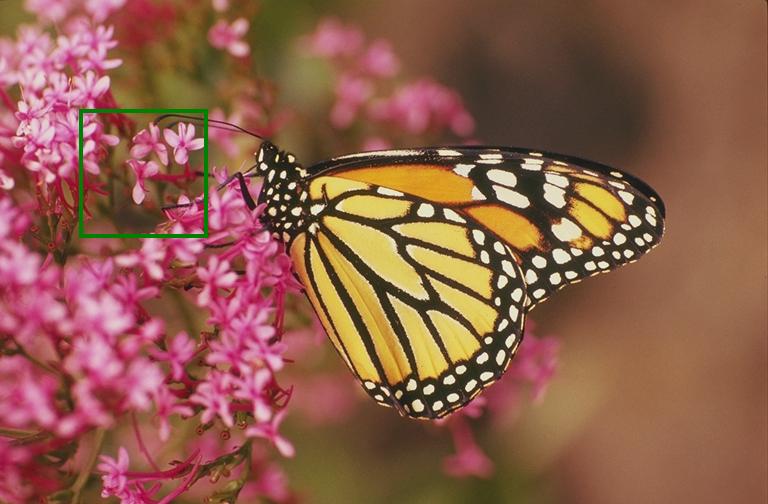}
            \caption*{Orig Image + Crop}
        \end{subfigure}
        \begin{subfigure}{0.20\textwidth}
            \includegraphics[width=\linewidth]{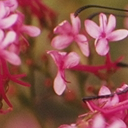}
            \caption*{HR\vspace*{1.1em}}
        \end{subfigure}            
        \begin{subfigure}{0.20\textwidth}
            \includegraphics[width=\linewidth]{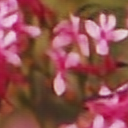}
            \caption*{SRGAN}
        \end{subfigure}       
        \begin{subfigure}{0.20\textwidth}
            \includegraphics[width=\linewidth]{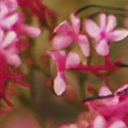}
            \caption*{ESRGAN}
        \end{subfigure}  
    \end{minipage}
    \,
    \begin{minipage}[t]{0.32\textwidth}
        \begin{subfigure}{0.35\linewidth}
            \includegraphics[width=\linewidth]{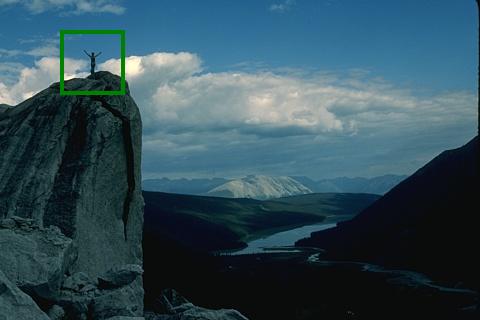}
            \caption*{Orig Image + Crop}
        \end{subfigure}
        \begin{subfigure}{0.20\textwidth}
            \includegraphics[width=\linewidth]{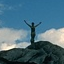}
            \caption*{HR\vspace*{1.1em}}
        \end{subfigure}            
        \begin{subfigure}{0.20\textwidth}
            \includegraphics[width=\linewidth]{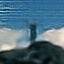}
            \caption*{SRGAN}
        \end{subfigure}        
        \begin{subfigure}{0.20\textwidth}
            \includegraphics[width=\linewidth]{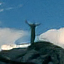}
            \caption*{ESRGAN}
        \end{subfigure}
    \end{minipage}
    \caption{Visual comparison of SRGAN vs ESRGAN without uncertainty estimation.
        ``HR'' indicates the $256\times 256$ high-resolution crop which is used as ground truth.
        ESRGAN looks qualitatively much more impressive than SRGAN:
        the latter's output is very blurry and seems unable to reconstruct fine-grained details.
        Conversely, the former is perceptually much closer to the original image and displays generally fewer artifacts.}
    \label{fig:visual_comparison_srgan_esrgan}
\end{figure*}

In this experiment we propose a visual comparison between our SRGAN and ESRGAN implementations, without applying uncertainty estimation. \Cref{fig:visual_comparison_srgan_esrgan} presents these results on three randomly selected images, showcasing that ESRGAN is still superior to SRGAN, with less artifacts and overall higher quality super-resolution results. For all future visual experiments, we will only show ESRGAN results, in particular for uncertainty estimation.

\subsection{Quantitative Uncertainty Analysis}

For SRGAN and ESRGAN, we compare the super-resolution performance after applying uncertainty estimation, we measure the peak signal to noise ratio (PSNR) and the structural similarity index (SSIM), as they are standard metrics for super-resolution, over the Set 5 and Set 14 datasets. Note that in this experiment we evaluate only the mean $\mu(\hat{Y})$, we evaluate uncertainty estimation performance in the coming experiments.

Table \ref{tab:uncert_num_comparison} shows our results. In both SR models, it is clear that Ensembles obtains the best performance in terms of PSRN, increasing task performance slightly on both datasets, but this is not always reflected on SSIM, as in some cases the baseline model without uncertainty estimation obtains a slightly better SSIM.

In Figure \ref{fig:ens_esrgan_visual}, we use error vs standard deviation plots to evaluate uncertainty quality of ESRGAN ensemble models. We built these plots by thresholding the model's standard deviation $\tilde{\sigma}(\hat{Y})$ from minimum to maximum value across a predicted SR output, and then computing the mean absolute error of the pixels passing the threshold. This measures how uncertainty predicts possible errors in the SR output and acts as a proxy for errors for each pixel.

All examples in Figure \ref{fig:ens_esrgan_visual} show the error increasing with standard deviation (uncertainty), indicating that model uncertainty is a reliable proxy for SR output errors. This results is visually confirmed in the same figure, as we additionally show per-pixel uncertainty maps, where higher uncertainty values visually correspond to SR results that are incorrect, like warped text, object boundaries, and small regions on which there is not enough information (pixels) to reconstruct correctly.

\subsection{Qualitative Uncertainty Analysis}

\begin{figure*}[t]
    \begin{subfigure}{0.19\textwidth}
        \includegraphics[width=\linewidth]{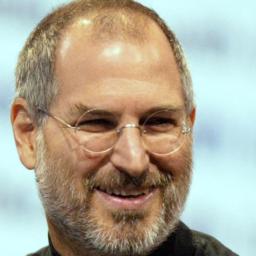}
        \caption{Ground truth}
    \end{subfigure}    
    \begin{subfigure}{0.19\textwidth}
        \includegraphics[width=\linewidth]{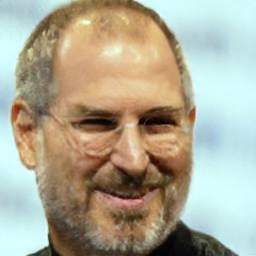}
        \caption{MCD SR}
    \end{subfigure}    
    \begin{subfigure}{0.19\textwidth}
        \includegraphics[width=\linewidth]{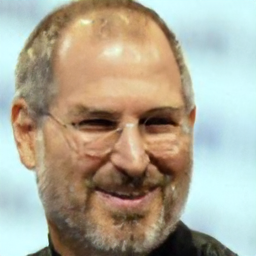}
        \caption{Ens SR}
    \end{subfigure}
    \begin{subfigure}{0.19\textwidth}
        \includegraphics[width=\linewidth]{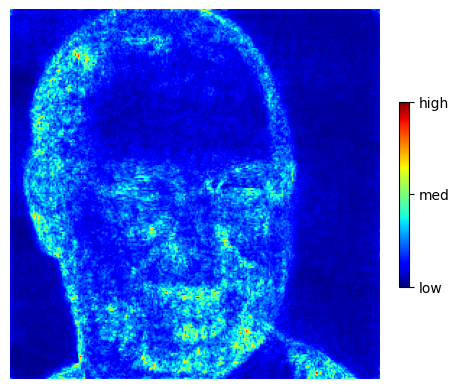} 
        \caption{MCD Uncert}
    \end{subfigure}
    \begin{subfigure}{0.19\textwidth}
        \includegraphics[width=\linewidth]{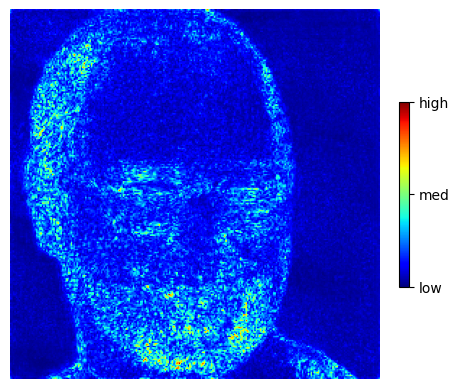} 
        \caption{Ens Uncert}
    \end{subfigure}
    \caption{Comparison of SR and its uncertainty between Ensembles and MC-Dropout.}
    \label{fig:uncert_visual_comparison_jobs}
\end{figure*}

\begin{figure*}[t]
    \centering
    \begin{subfigure}{0.15\textwidth}
        \includegraphics[width=\linewidth]{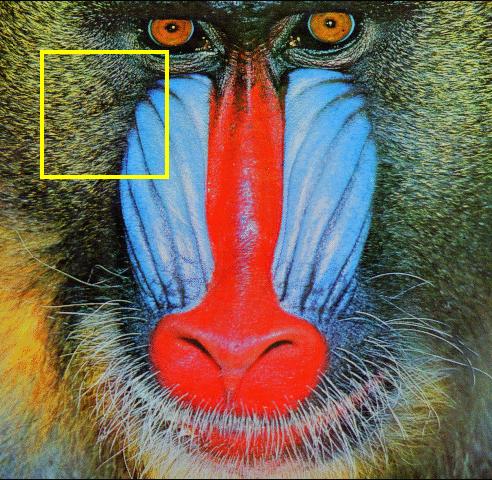}
        \caption{Baboon\vspace*{1.15em}}
    \end{subfigure}    
    \begin{subfigure}{0.15\textwidth}
        \includegraphics[width=\linewidth]{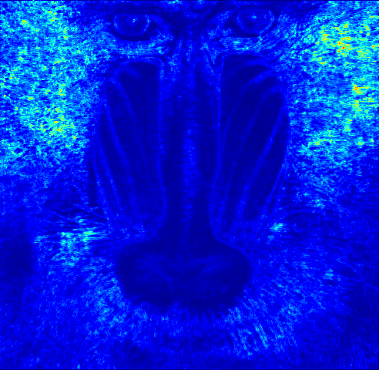}
        \caption{MCD\vspace*{1.15em}}
    \end{subfigure}
    \begin{subfigure}{0.15\textwidth}
        \includegraphics[width=\linewidth]{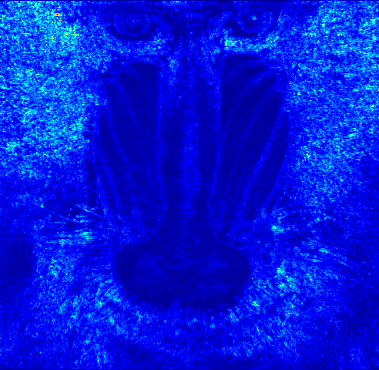} 
        \caption{Ensembles\vspace*{1.15em}}
    \end{subfigure}    
    \begin{subfigure}{0.15\textwidth}
        \includegraphics[width=\linewidth]{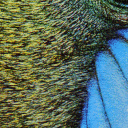}
        \caption{Crop\\(PSNR/SSIM)}
    \end{subfigure}    
    \begin{subfigure}{0.15\textwidth}
        \includegraphics[width=\linewidth]{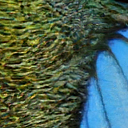}
        \caption{MCD\\(15.90/0.31)}
    \end{subfigure}
    \begin{subfigure}{0.15\textwidth}
        \includegraphics[width=\linewidth]{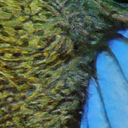} 
        \caption{Ensembles\\(17.72/0.36)}
    \end{subfigure}
    \caption{Visual comparison of Super-Resolution output and Uncertainty maps for baboon.png in Set14.}
    \label{fig:uncert_visual_comparison_baboon}
\end{figure*}

\begin{figure*}[t]
    \centering
    \begin{subfigure}{0.18\textwidth}
        \includegraphics[width=\linewidth]{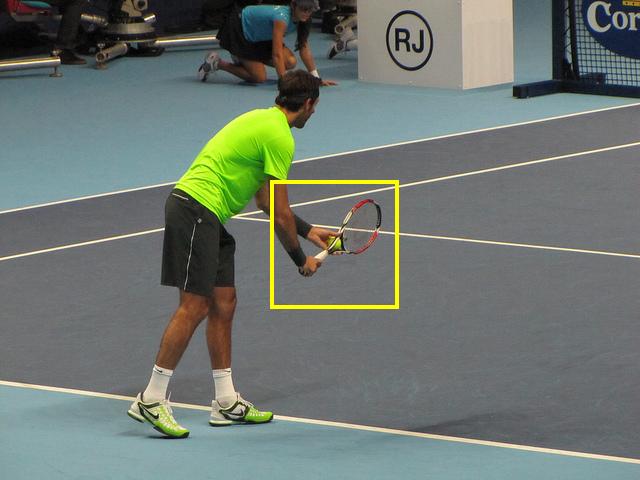}
        \caption{COCO Image\vspace*{1.15em}}
    \end{subfigure}    
    \begin{subfigure}{0.18\textwidth}
        \includegraphics[width=\linewidth]{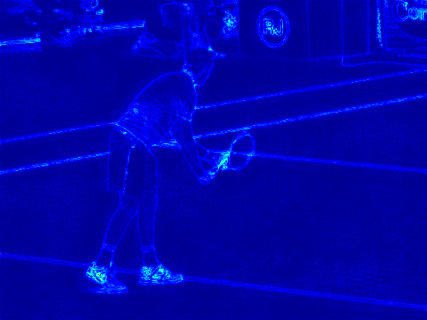}
        \caption{MCD\vspace*{1.15em}}
    \end{subfigure}
    \begin{subfigure}{0.18\textwidth}
        \includegraphics[width=\linewidth]{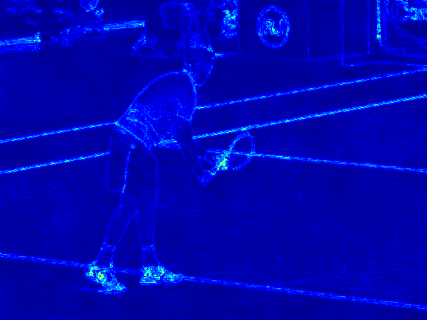} 
        \caption{Ensembles\vspace*{1.15em}}
    \end{subfigure}    
    \begin{subfigure}{0.12\textwidth}
        \includegraphics[width=\linewidth]{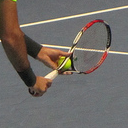}
        \caption{Crop\\(PSNR/SSIM)}
    \end{subfigure}    
    \begin{subfigure}{0.12\textwidth}
        \includegraphics[width=\linewidth]{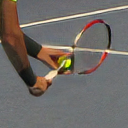}
        \caption{MCD\\(26.43/0.81)}
    \end{subfigure}
    \begin{subfigure}{0.12\textwidth}
        \includegraphics[width=\linewidth]{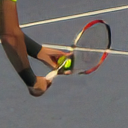} 
        \caption{Ensembles\\(27.14/0.84)}
    \end{subfigure}
    \caption{Visual comparison of Super-Resolution output and Uncertainty maps for COCO Tennis image.}
    \label{fig:uncert_visual_comparison_tennis}
\end{figure*}

This experiments makes a qualitative analysis of ESRGAN, comparing MC-Dropout and Ensembles. Figures \ref{fig:uncert_visual_comparison_jobs}, \ref{fig:uncert_visual_comparison_baboon}, and \ref{fig:uncert_visual_comparison_tennis} display these results.

Figure \ref{fig:uncert_visual_comparison_jobs} is particularly challenging, as upscaling the scalp hair and beard is very difficult as it is fine detail that is not fully present in the low-resolution input, and both MC-Dropout and Ensembles indicate higher uncertainty in the scalp hair and beard areas, corresponding to more erroneous predictions.

Figures \ref{fig:uncert_visual_comparison_baboon} and \ref{fig:uncert_visual_comparison_tennis} show full size uncertainty maps and crops detailing high uncertainty regions, showing how upscaling made by MC-Dropout and Ensembles differs, in particular ensembles seems to produce slightly blurrier regions, but the focus crops correspond to regions that are very hard to upscale (like Baboon hair or Tennis Racket and Ball overlap), including high frequency details that cannot be upscaled correctly given a low-resolution image.

\subsection{SR Error Detection Examples}

\begin{figure*}[t]
    \centering
    \begin{minipage}{0.48\textwidth}
        \begin{subfigure}{0.45\linewidth}
            \includegraphics[width=\linewidth]{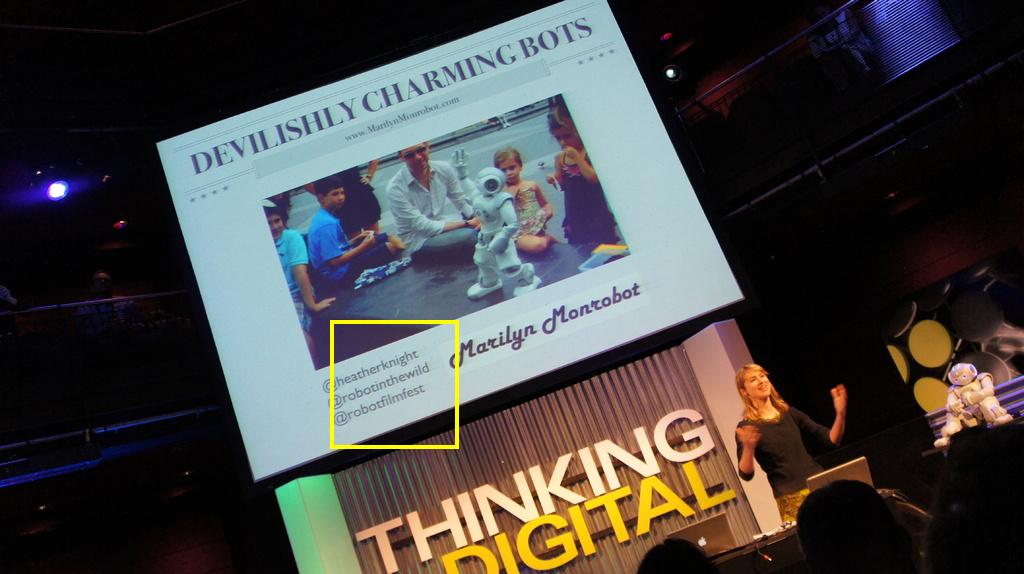}
            \caption{Image}
        \end{subfigure}    
        \begin{subfigure}{0.45\linewidth}
            \includegraphics[width=\linewidth]{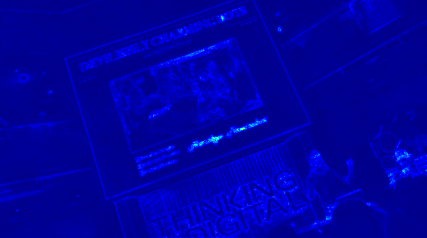}
            \caption{SR}
        \end{subfigure}
        
        \vspace*{0.5cm}

        \begin{subfigure}{0.30\linewidth}
            \includegraphics[width=\linewidth]{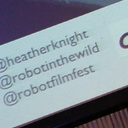}
            \caption{Crop}
        \end{subfigure}    
        \begin{subfigure}{0.30\linewidth}
            \includegraphics[width=\linewidth]{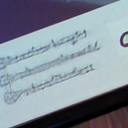}
            \caption{SR}
        \end{subfigure}
        \begin{subfigure}{0.30\linewidth}
            \includegraphics[width=\linewidth]{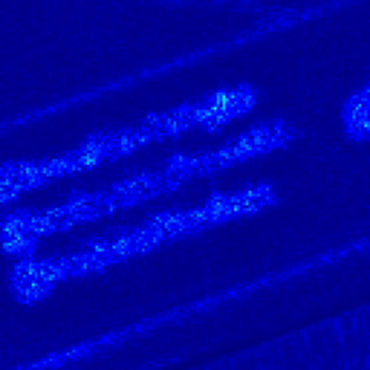} 
            \caption{Unc Map}
        \end{subfigure}
    \end{minipage}    
    \begin{minipage}{0.48\textwidth}
        \begin{subfigure}{0.45\linewidth}
            \includegraphics[width=\linewidth]{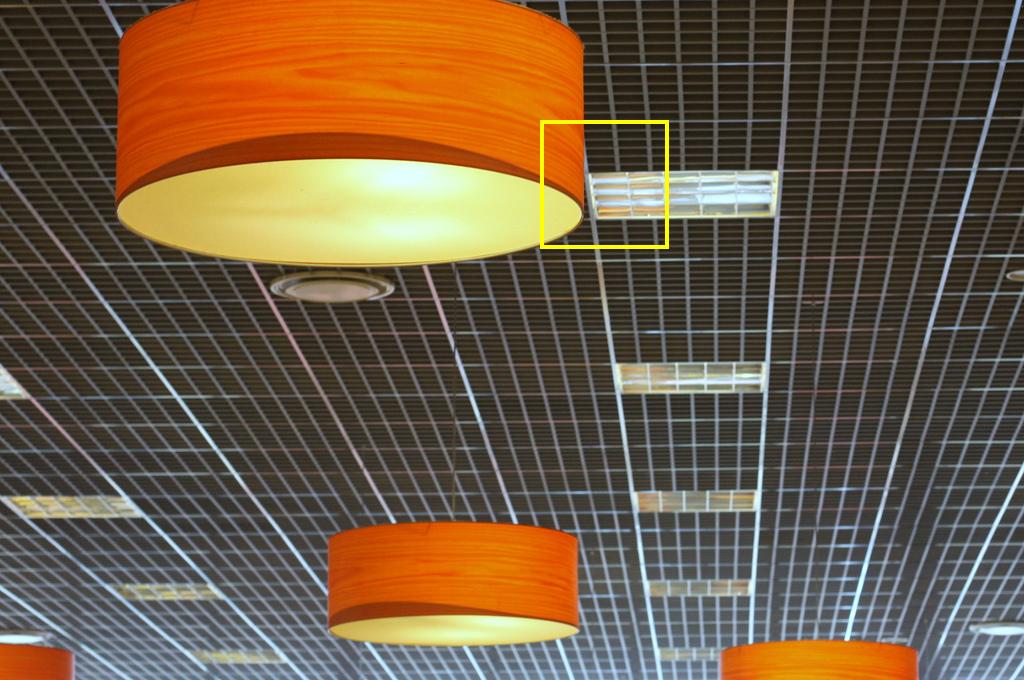}
            \caption{Image}
        \end{subfigure}    
        \begin{subfigure}{0.45\linewidth}
            \includegraphics[width=\linewidth]{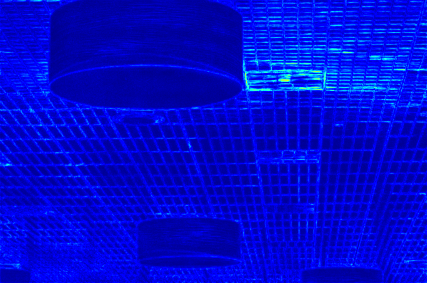}
            \caption{SR}
        \end{subfigure}
        
        \vspace*{0.5cm}

        \begin{subfigure}{0.30\linewidth}
            \includegraphics[width=\linewidth]{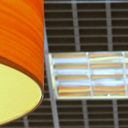}
            \caption{Crop}
        \end{subfigure}    
        \begin{subfigure}{0.30\linewidth}
            \includegraphics[width=\linewidth]{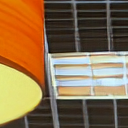}
            \caption{SR}
        \end{subfigure}
        \begin{subfigure}{0.30\linewidth}
            \includegraphics[width=\linewidth]{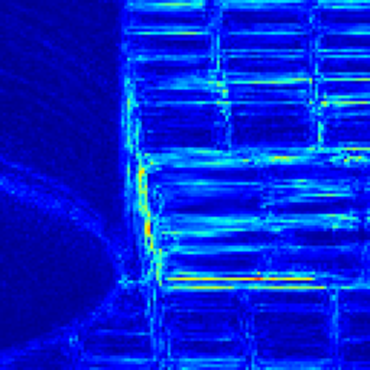} 
            \caption{Unc Map}
        \end{subfigure}
    \end{minipage}
    \caption{Two examples of uncertainty pointing to improper reconstructions/errors.}
    \label{fig:visual_uncertainty_errors}
\end{figure*}

Finally, in \Cref{fig:visual_uncertainty_errors}, we showcase some selected examples where output uncertainty maps are particularly useful to detect erroneous upscaling results. In particular the SR algorithms struggle with fine details like text and high frequency regular patterns. These results complement our previous findings, from which there is a clear conclusion: uncertainty maps produced by Ensembles can provide additional information to a human user, to determine which SR regions are reliable (low pixel error) and which ones are not (high pixel error), and uncertainty maps can be used as additional information for further use of a SR result.

\section{\uppercase{Conclusions and Future Work}}

In the present paper we built SRGAN and ESRGAN models with uncertainty estimation for super-resolution, using MC-Dropout and Ensembles.
The aim was to detect SR output regions in which these models are more uncertain, indicating that they might correspond to incorrect upscaling outputs. We extensively validated our proposed approach on several datasets and over multiple facets, including a qualitative analysis of SR outputs and uncertainty maps, and quantitative metrics like error vs standard deviation plots.

Overall, we believe our results show that uncertainty estimation has good potential for super-resolution applications, as human users can use uncertainty maps together with the SR output to decide if they should trust the SR image in a region-by-region basis, as uncertainty is a proxy for super-resolution correctness.

\paragraph{Limitations} Our work is limited by the selection of uncertanity estimation methods (MC-Dropout and Ensembles), and for datasets we used for training and evaluation. Our aim was not to build the most precise SR model, but to evaluate the possibilities of building SR models with uncertainty estimation.

\paragraph{Broader Societal Impact} SR for images and video has a especial place in the public due to series like Crime Scene Investigation (CSI) that popularized magical thinking about super-resolution \cite{allen2007reading}; SR models are, however, imperfect and cannot correctly upscale every possible input, especially when there is a large amount of missing information. We expect that SR models with uncertainty can signal to the user when the SR outputs are not reliable, improving societal understanding of these methods and directly indicating that models can make mistakes and should not be trusted blindly.

\bibliographystyle{apalike}
{\small
\bibliography{biblio}}

\begin{thebibliography}{}

\bibitem[Agustsson and Timofte, 2017]{agustsson2017ntire}
Agustsson, E. and Timofte, R. (2017).
\newblock Ntire 2017 challenge on single image super-resolution: Dataset and
  study.
\newblock In {\em Proceedings of the IEEE conference on computer vision and
  pattern recognition workshops}, pages 126--135.

\bibitem[Allen, 2007]{allen2007reading}
Allen, M. (2007).
\newblock {\em Reading'CSI': Crime TV Under the Microscope}.
\newblock Bloomsbury Publishing.

\bibitem[Bevilacqua et~al., 2012]{bevilacqua2012low}
Bevilacqua, M., Roumy, A., Guillemot, C., and Alberi-Morel, M.~L. (2012).
\newblock Low-complexity single-image super-resolution based on nonnegative
  neighbor embedding.

\bibitem[Deng et~al., 2009]{deng2009imagenet}
Deng, J., Dong, W., Socher, R., Li, L.-J., Li, K., and Fei-Fei, L. (2009).
\newblock Imagenet: A large-scale hierarchical image database.
\newblock In {\em 2009 IEEE conference on computer vision and pattern
  recognition}, pages 248--255. Ieee.

\bibitem[Gal and Ghahramani, 2016]{gal2016dropout}
Gal, Y. and Ghahramani, Z. (2016).
\newblock Dropout as a bayesian approximation: Representing model uncertainty
  in deep learning.
\newblock In {\em international conference on machine learning}, pages
  1050--1059. PMLR.

\bibitem[Goodfellow et~al., 2014]{goodfellow2014generative}
Goodfellow, I., Pouget-Abadie, J., Mirza, M., Xu, B., Warde-Farley, D., Ozair,
  S., Courville, A., and Bengio, Y. (2014).
\newblock Generative adversarial nets.
\newblock {\em Advances in neural information processing systems}, 27.

\bibitem[He et~al., 2015]{he2015delving}
He, K., Zhang, X., Ren, S., and Sun, J. (2015).
\newblock Delving deep into rectifiers: Surpassing human-level performance on
  imagenet classification.
\newblock In {\em Proceedings of the IEEE international conference on computer
  vision}, pages 1026--1034.

\bibitem[Huang et~al., 2015]{huang2015single}
Huang, J.-B., Singh, A., and Ahuja, N. (2015).
\newblock Single image super-resolution from transformed self-exemplars.
\newblock In {\em Proceedings of the IEEE conference on computer vision and
  pattern recognition}, pages 5197--5206.

\bibitem[Jolicoeur-Martineau, 2018]{jolicoeur2018relativistic}
Jolicoeur-Martineau, A. (2018).
\newblock The relativistic discriminator: a key element missing from standard
  gan.
\newblock {\em arXiv preprint arXiv:1807.00734}.

\bibitem[Kar and Biswas, 2021]{kar2021fast}
Kar, A. and Biswas, P.~K. (2021).
\newblock Fast bayesian uncertainty estimation and reduction of batch
  normalized single image super-resolution network.
\newblock In {\em Proceedings of the IEEE/CVF Conference on Computer Vision and
  Pattern Recognition}, pages 4957--4966.

\bibitem[Kingma and Ba, 2014]{kingma2014adam}
Kingma, D.~P. and Ba, J. (2014).
\newblock Adam: A method for stochastic optimization.
\newblock {\em arXiv preprint arXiv:1412.6980}.

\bibitem[Lakshminarayanan et~al., 2017]{lakshminarayanan2017simple}
Lakshminarayanan, B., Pritzel, A., and Blundell, C. (2017).
\newblock Simple and scalable predictive uncertainty estimation using deep
  ensembles.
\newblock {\em Advances in neural information processing systems}, 30.

\bibitem[Ledig et~al., 2017]{ledig2017photo}
Ledig, C., Theis, L., Husz{\'a}r, F., Caballero, J., Cunningham, A., Acosta,
  A., Aitken, A., Tejani, A., Totz, J., Wang, Z., et~al. (2017).
\newblock Photo-realistic single image super-resolution using a generative
  adversarial network.
\newblock In {\em Proceedings of the IEEE conference on computer vision and
  pattern recognition}, pages 4681--4690.

\bibitem[Lin et~al., 2014]{lin2014microsoft}
Lin, T.-Y., Maire, M., Belongie, S., Hays, J., Perona, P., Ramanan, D.,
  Doll{\'a}r, P., and Zitnick, C.~L. (2014).
\newblock Microsoft coco: Common objects in context.
\newblock In {\em Computer Vision--ECCV 2014: 13th European Conference, Zurich,
  Switzerland, September 6-12, 2014, Proceedings, Part V 13}, pages 740--755.
  Springer.

\bibitem[Liu et~al., 2023]{liu2023spectral}
Liu, T., Cheng, J., and Tan, S. (2023).
\newblock Spectral bayesian uncertainty for image super-resolution.
\newblock In {\em Proceedings of the IEEE/CVF Conference on Computer Vision and
  Pattern Recognition}, pages 18166--18175.

\bibitem[Martin et~al., 2001]{martin2001database}
Martin, D., Fowlkes, C., Tal, D., and Malik, J. (2001).
\newblock A database of human segmented natural images and its application to
  evaluating segmentation algorithms and measuring ecological statistics.
\newblock In {\em Proceedings Eighth IEEE International Conference on Computer
  Vision. ICCV 2001}, volume~2, pages 416--423. IEEE.

\bibitem[Shi et~al., 2016]{shi2016real}
Shi, W., Caballero, J., Husz{\'a}r, F., Totz, J., Aitken, A.~P., Bishop, R.,
  Rueckert, D., and Wang, Z. (2016).
\newblock Real-time single image and video super-resolution using an efficient
  sub-pixel convolutional neural network.
\newblock In {\em Proceedings of the IEEE Conference on Computer Vision and
  Pattern Recognition (CVPR)}, pages 1874--1883.

\bibitem[Simonyan and Zisserman, 2014]{simonyan2014very}
Simonyan, K. and Zisserman, A. (2014).
\newblock Very deep convolutional networks for large-scale image recognition.
\newblock {\em arXiv preprint arXiv:1409.1556}.

\bibitem[Song and Yang, 2023]{song2023uncertainty}
Song, H. and Yang, Y. (2023).
\newblock Uncertainty quantification in super-resolution guided wave array
  imaging using a variational bayesian deep learning approach.
\newblock {\em NDT \& E International}, 133:102753.

\bibitem[Srivastava et~al., 2014]{srivastava2014dropout}
Srivastava, N., Hinton, G., Krizhevsky, A., Sutskever, I., and Salakhutdinov,
  R. (2014).
\newblock Dropout: a simple way to prevent neural networks from overfitting.
\newblock {\em The journal of machine learning research}, 15(1):1929--1958.

\bibitem[Tanno et~al., 2017]{tanno2017bayesian}
Tanno, R., Worrall, D.~E., Ghosh, A., Kaden, E., Sotiropoulos, S.~N.,
  Criminisi, A., and Alexander, D.~C. (2017).
\newblock Bayesian image quality transfer with cnns: exploring uncertainty in
  dmri super-resolution.
\newblock In {\em Medical Image Computing and Computer Assisted Intervention-
  MICCAI 2017: 20th International Conference, Quebec City, QC, Canada,
  September 11-13, 2017, Proceedings, Part I 20}, pages 611--619. Springer.

\bibitem[Timofte et~al., 2017]{timofte2017ntire}
Timofte, R., Agustsson, E., Van~Gool, L., Yang, M.-H., and Zhang, L. (2017).
\newblock Ntire 2017 challenge on single image super-resolution: Methods and
  results.
\newblock In {\em Proceedings of the IEEE conference on computer vision and
  pattern recognition workshops}, pages 114--125.

\bibitem[Wang et~al., 2018]{wang2018esrgan}
Wang, X., Yu, K., Wu, S., Gu, J., Liu, Y., Dong, C., Qiao, Y., and Change~Loy,
  C. (2018).
\newblock Esrgan: Enhanced super-resolution generative adversarial networks.
\newblock In {\em Proceedings of the European conference on computer vision
  (ECCV) workshops}, pages 0--0.

\bibitem[Zeyde et~al., 2012]{zeyde2012single}
Zeyde, R., Elad, M., and Protter, M. (2012).
\newblock On single image scale-up using sparse-representations.
\newblock In {\em Curves and Surfaces: 7th International Conference, Avignon,
  France, June 24-30, 2010, Revised Selected Papers 7}, pages 711--730.
  Springer.

\bibitem[Zhang et~al., 2021]{zhang2021benchmarking}
Zhang, K., Li, D., Luo, W., Ren, W., Stenger, B., Liu, W., Li, H., and Yang,
  M.-H. (2021).
\newblock Benchmarking ultra-high-definition image super-resolution.
\newblock In {\em Proceedings of the IEEE/CVF international conference on
  computer vision}, pages 14769--14778.

\end{thebibliography}

\end{document}